\documentstyle[12pt]{article}

\begin{document}

\noindent {\small CITUSC/00-011\hfill \hfill hep-th/0002140 \newline
}

{\small \hfill }

{\vskip1cm}

\begin{center}
{\Large {\bf Two-Time Physics with Gravitational}}\\[0pt]
{\Large {\bf and Gauge Field Backgrounds}}{\footnote{%
This research was partially supported by the US. Department of Energy under
grant number DE-FG03-84ER40168.}}

\bigskip

{\vskip0.5cm}

{\bf Itzhak Bars}

{\vskip1cm}

{Department of Physics and Astronomy}

and

CIT-USC Center for Theoretical Physics

{University of Southern California}

{\ Los Angeles, CA 90089-0484, USA}

{\vskip1cm}

{\bf Abstract}

{\vskip0.5cm}
\end{center}

It is shown that all possible gravitational, gauge and other interactions
experienced by particles in ordinary $d$-dimensions (one-time) can be
described in the language of two-time physics in a spacetime with $d+2$
dimensions. This is obtained by generalizing the worldline formulation of
two-time physics by including background fields. A given two-time model,
with a fixed set of background fields, can be gauged fixed from $d+2$
dimensions to $\left( d-1\right) +1$ dimensions to produce diverse one-time
dynamical models, all of which are dually related to each other under the
underlying gauge symmetry of the unified two-time theory. To satisfy the
gauge symmetry of the two-time theory the background fields must obey
certain coupled differential equations that are generally covariant and
gauge invariant in the target $d+2$ dimensional spacetime. The gravitational
background obeys a closed homothety condition while the gauge field obeys a
differential equation that generalizes a similar equation derived by Dirac
in 1936. Explicit solutions to these coupled equations show that the usual
gravitational, gauge, and other interactions in $d$ dimensions may be viewed
as embedded in the higher $d+2$ dimensional space, thus displaying higher
spacetime symmetries that otherwise remain hidden.

\newpage

\section{Introduction}

Two-Time Physics \cite{conf}-\cite{string2t} is an approach that provides a
new perspective for understanding ordinary one-time dynamics from a higher
dimensional, more unified point of view including two timelike dimensions.
This is achieved by introducing new gauge symmetries that insure unitarity,
causality and absence of ghosts. The new phenomenon in two-time physics is
that the gauge symmetry can be used to obtain various one-time dynamical
systems from the same simple action of two-time physics, through gauge
fixing, thus uncovering a new layer of unification through higher dimensions.

The principle behind two-time physics is the gauge symmetry \cite{conf}. The
basic observation in its simplest form is that for any theory the Lagrangian
has the form $L=\frac{1}{2}\left( \dot{x}p-\dot{p}x\right) -H\left(
x,p\right) $ up to an inessential total time derivative. The first term has
a global Sp$\left( 2,R\right) $ symmetry that transforms $\left( x,p\right) $
as a doublet. The basic question we pose is: what modification of the
Lagrangian can turn this global symmetry into a local symmetry? The reason
to be interested in such a local symmetry is that duality symmetries in
M-theory and N=2 super Yang-Mills theory have similarities to gauge
symplectic transformations, and their origin in the fundamental theories in
physics remains a mystery. Understanding them may well be the key to
constructing M-theory. Independent of M-theory, the question is a
fundamental one in its own right, and its investigation has already led to a
reformulation of ordinary one-time dynamical systems in a new language of
two-time physics. This has uncovered previously unnoticed higher symmetries
in well known one-time dynamical systems, and provided a new level of
unification through higher dimensions for systems that previously would have
been considered unrelated to each other \cite{lifting}. The simplest Sp$%
\left( 2,R\right) $ gauge symmetry has generalizations when spin \cite{spin}%
, supersymmetry \cite{super2t}\cite{liftM}, and extended objects (branes) 
\cite{string2t} are part of the theory. Recent works have given an
indication that the domain of unification of two-time physics can be
enlarged in additional directions in field theory \cite{fieldth} including
interactions, and in the world of branes \cite{gibbons-etal}.

In the two-time physics approach the familiar one-time is a gauge dependent
concept. From the point of view of a two-time observer the true gauge
invariants are identical in a variety of one-time dynamical systems that are
unified by the same two-time action. Such gauge invariant quantities can be
used to test the validity of the underlying unification. An important gauge
invariant concept is the global symmetry of the two-time action, which must
be shared by all the gauge fixed one-time dynamical systems. In the simplest
case the global symmetry is SO$\left( d,2\right) $, but this can be
different in the presence of background fields as we will see in the current
paper. In the simple case, the SO$\left( d,2\right) $ symmetry has been
shown to be present {\it in the same irreducible representation} in all the
one-time dynamical systems derived from the same two-time action. The
presence of such symmetries, which remained unknown even in elementary
one-time systems until the advent of two-time physics, can be considered as
a test of the underlying unification within a two-time theory \cite{lifting}.

Two-Time Physics has been generalized to include global space-time
supersymmetry and local kappa supersymmetry with two-times \cite{super2t}.
This led to a framework which suggests that M-theory could be embedded in a
two-time theory in 13 dimensions, with a global OSp(1$|$64) symmetry. In
this scenario the different corners of M-theory correspond to gauge fixed
sectors of the 13D theory, and the dualities in M-theory are regarded as
gauge transformations from one fixed gauge to another fixed gauge. Then the
well known supersymmetries of various corners of M-theory appear as
subsupergroups of OSp$\left( 1|64\right) $. This mechanism has been
illustrated through explicit examples of dynamical particle models \cite
{liftM}\cite{toyM} which may be regarded as a toy-M-theory. In the
11D-covariant gauge fixed corner, the supergroup OSp(1$|$64) is interpreted
as the conformal supergroup in 11-dimensions, with 32 supersymmetries and 32
superconformal symmetries. But in other gauge fixed sectors, the same OSp(1$%
| $64) symmetry of two-time physics is realized and interpreted differently,
thus revealing various corners of toy-M-theory on which a subsupergroup is
linearly realized while the rest is non-linearly realized. Indeed OSp(1$|$%
64) contains various embeddings that reveal 13,12,11 dimensional
supersymmetries, as well as the usual 10-dimensional type-IIA, type-IIB,
heterotic, type-I, and AdS$_{D}\otimes $S$^{k}$ type supersymmetries in $%
D+k=11,10$ and lower dimensions. The explicit models provided by \cite{liftM}
\cite{toyM} illustrate these ideas while beginning to realize dynamically
some of the observations that suggested two-time physics in the framework of
branes, dualities and extended supersymmetries in M-theory, F-theory, and
S-theory \cite{duff}-\cite{nishino}.

In this paper we generalize the worldline formulation of two-time physics by
including background gravitational and gauge fields and other potentials. To
keep the discussion simple we concentrate mainly on particles without
supersymmetry. For spinless particles, as in the case of the free theory,
local Sp($2,R$) gauge symmetry is imposed as the underlying principle. For
the gauge symmetry to be valid, the gravitational and gauge fields and other
potentials must obey certain differential equations. We show that the gauge
field obeys an equation that generalizes a similar one discovered by Dirac
in 1936 \cite{Dirac} in the flat background, while the gravitational field
satisfies a closed homothety condition. When all fields are simultaneously
present they obey coupled equations. Examples of background fields that
solve these equations are provided.

A similar treatment for spinning particles in background fields is given. As
in the free theory, local OSp$\left( n|2\right) $ gauge symmetry is imposed
as the underlying principle. The set of background fields is now richer. The
generalizations of Dirac's equation and the closed homothety conditions in
the presence of spin are derived. Instead of OSp$\left( n|2\right) $ gauge
symmetry it may also be possible to consider other supergroups that contain
Sp$\left( 2,R\right) \equiv SL\left( 2,R\right) .$

In the presence of the background fields one learns that much larger classes
of one-time dynamical systems can now be reformulated as gauge fixed
versions of the same two-time theory. This extends the domain of unification
of one-time systems through higher dimensions and a sort of duality symmetry
(the Sp$(2,R)$ gauge symmetry and its generalizations in systems with spin
and/or spacetime supersymmetry, and branes). Furthermore, with the results
of this paper it becomes evident that all one-time particle dynamics can be
reformulated as particle dynamics in two-time physics. This provides a much
broader realm of possible applications of the two-time physics formalism.

One possible practical application of the formulation is to provide a tool
for solving problems by transforming a complicated one-time dynamical system
(one fixed gauge) to a simpler one-time dynamical system (another fixed
gauge), as in duality transformations in M-theory. Although this may turn
out to be the computationally useful aspect of this formulation, it is not
explored in the present paper since our main aim here is the formulation of
the concepts.

The two-time formulation also has deeper ramifications. By providing the
perspective of two-time physics for ordinary physical phenomena, the
familiar ``time'' dimension appears to play a less fundamental role in the
formulation of physics. Since the usual ``time'' is a gauge dependent
concept in the new formulation, naturally one is led to a re-examination of
the concept of ``time'' in this new setting.

\section{Local and global symmetry}

We start with a brief summary of the worldline formulation of two-time
physics for the simplest case of spinless particle dynamics without
background fields and without a Hamiltonian \cite{conf}-\cite{liftM} (i.e.
the ``free'' case). Just demanding local symmetry for the first term in the
Lagrangian gives a surprisingly rich model based on $Sp(2,R)$ gauge symmetry
described by the action 
\begin{eqnarray}
S_{0} &=&\frac{1}{2}\int d\tau \,D_{\tau }X_{i}^{M}X_{j}^{N}\varepsilon
^{ij}\eta _{MN}  \label{action} \\
&=&\int d\tau (\partial _{\tau }X_{1}^{M}X_{2}^{N}-\frac{1}{2}%
A^{ij}X_{i}^{M}X_{j}^{N})\eta _{MN}.  \label{confor}
\end{eqnarray}
Here $X_{i}^{M}(\tau )$ is an $Sp(2,R)$ doublet, consisting of a coordinate
and its conjugate momentum ($X_{1}^{M}\equiv X^{M}$ and $X_{2}^{M}\equiv
P^{M}$). The indices $i,j=1,2$ denote the doublet of $Sp(2,R)$; they are
raised and lowered by the antisymmetric Levi-Civita symbol $\varepsilon
_{ij} $. The covariant derivative $D_{\tau }X_{i}^{M}$ that appears in (\ref
{action}) is defined as 
\begin{equation}
D_{\tau }X_{i}^{M}=\partial _{\tau }X_{i}^{M}-\varepsilon
_{ik}A^{kl}X_{l}^{M},
\end{equation}
where $A^{ij}\left( \tau \right) $ are the three Sp$\left( 2,R\right) $
gauge potentials in the adjoint representation written as a 2$\times 2$
symmetric matrix. The local $Sp(2,R)$ acts as $\delta X_{i}^{M}=\varepsilon
_{ik}\omega ^{kl}X_{l}^{m}$ and $\delta A^{ij}=\omega ^{ik}\varepsilon
_{kl}A^{kj}+\omega ^{jk}\varepsilon _{kl}A^{ik}+\partial _{\tau }\omega
^{ij} $, where $\omega ^{ij}\left( \tau \right) $ are the Sp$\left(
2,R\right) $ gauge parameters. The second form of the action (\ref{confor})
is obtained after an integration by parts so that only $X_{1}^{M}$ appears
with derivatives. This allows the identification of $X,P$ by the canonical
procedure ($X_{1}^{M}\equiv X^{M}$ and $X_{2}^{M}\equiv P^{M}=\partial
S_{0}/\partial \dot{X}_{1M}$). A third form of the action can be obtained by
integrating out $X_{2}^{M}$ and writing it in terms of $X^{M}$ and $\dot{X}%
^{M}$ \cite{conf}\cite{marnelius}. Then the local Sp$\left( 2,R\right) =$SO$%
\left( 1,2\right) $ can also be regarded as the local conformal group on the
worldline (including $\tau $ reparametrization, local scale transformations,
and local special conformal transformations) and the theory can be
interpreted as conformal gravity on the worldline \cite{conf}\cite{string2t}.

The gauge fields $A^{11}$, $A^{12}$, and $A^{22}$ act as Lagrange
multipliers for the following three first class constraints 
\begin{equation}
Q_{ij}^{0}=X_{i}\cdot X_{j}=0\,\,\rightarrow \,\,\,X^{2}=P^{2}=X\cdot P=0,
\end{equation}
as implied by the local $Sp(2,R)$ invariance. From the basic quantum rules
for $\left( X^{M},P^{M}\right) $ one can verify that the $Q_{ij}^{0}$ form
the Sp$\left( 2,R\right) $ algebra 
\begin{eqnarray}
\left[ Q_{ij},Q_{kl}\right] &=&i\varepsilon _{jk}Q_{il}+i\varepsilon
_{ik}Q_{jl}+i\varepsilon _{jl}Q_{ik}+i\varepsilon _{il}Q_{jk}\,,\quad or
\label{sp2alg} \\
\left[ Q_{11},Q_{22}\right] &=&4iQ_{12},\quad \left[ Q_{11},Q_{12}\right]
=2iQ_{11},\quad \left[ Q_{22},Q_{12}\right] =-2iQ_{22}.
\end{eqnarray}

The two timelike dimensions are not put in by hand, they are implied by the
local $Sp(2,R)$ symmetry. It is precisely the solution of the constraints $%
Q_{ij}^{0}=0$ that require the global metric $\eta _{MN}$ in (\ref{action})
to have a signature with two-time like dimensions: if $\eta _{MN}$ were
purely Euclidean the only solution would be vanishing vectors $X_{i}^{M}$,
if it had Minkowski signature (one time) the only solution would be two
lightlike parallel vectors $X_{i}^{M}$ without any angular momentum, if it
had more than two timelike dimensions there would be ghosts that would
render the theory non-unitary. The local Sp$\left( 2,R\right) $ is just
enough gauge symmetry to remove the ghosts due to two timelike dimensions.
Thus, $\eta _{MN}$ stands for the flat metric on a ($d,2$) dimensional
space-time. It is the only signature consistent with absence of ghosts,
unitrarity or causality problems.

We now turn to the global symmetries that are gauge invariant under Sp$%
\left( 2,R\right) $. The metric $\eta _{MN}$ is invariant under SO$\left(
d,2\right) .$ Hence the action (\ref{action}-\ref{confor}) has an explicit
global $SO(d,2)$ invariance. Like the two times, the $SO(d,2)$ symmetry of
the action (\ref{action}) is also implied by the local $Sp(2,R)$ symmetry
when background fields are absent. The SO$\left( d,2\right) $ Lorentz
generators 
\begin{equation}
L^{MN}=X^{M}P^{N}-X^{N}P^{M}=\varepsilon ^{ij}X_{i}^{M}X_{j}^{N}  \label{lmn}
\end{equation}
commute with the Sp$\left( 2,R\right) $ generators, therefore they are gauge
invariant. As we mentioned above, different gauge choices lead to different
one-time particle dynamics (examples: free massless and massive particles,
H-atom, harmonic oscillator, particle in AdS$_{D}\times $S$^{k}$ etc.) all
of which have $SO(d,2)$ {\it invariant actions} that are directly obtained
from (\ref{action}-\ref{confor}) by gauge fixing. Since the action (\ref
{action}-\ref{confor}) and the generators $L^{MN}$ are gauge invariant, the
global symmetry SO$\left( d,2\right) $ is not lost by gauge fixing. This
explains why one should expect a hidden (previously unnoticed, non-linearly
realized) global symmetry SO$\left( d,2\right) $ for each of the one-time
systems that result by gauge fixing\footnote{%
A well known case is the SO$\left( 4,2\right) $ conformal symmetry of the
massless particle. Less well known is the SO$\left( 4,2\right) $ symmety of
the H-atom action, which acts as the dynamical symmetry for the quantum
H-atom. Previously unknown is the SO$\left( 4,2\right) $ symmetry of the
massive non-relativistic particle action $S=\int d\tau $ ${\bf \dot{x}}%
^{2}\,/2m$. Others are the SO$\left( 10,2\right) $ symmetry of a particle in
the AdS$_{5}\times S^{5}$ background, or the SO$\left( 11,2\right) $
symmetry in the AdS$_{7}\times S^{4}$ and the AdS$_{4}\times S^{7}$
backgrounds, etc. These and more examples of such non-linearly realized SO$%
\left( d,2\right) $ hidden symmetries for familiar systems in any space-time
dimension $d$ are explicitly given in \cite{lifting}.}. Furthermore all of
the resulting one-time dynamical systems are quantum mechanically realized
in the {\it same unitary representation} of SO$\left( d,2\right) $ \cite
{conf}-\cite{lifting}. This fact can be understood again as a simple
consequence of representing the same quantum mechanical two-time system in
various fixed gauges. The gauge choices merely distinguish one basis versus
another basis within the same unitary representation of SO$\left( d,2\right) 
$ without changing the Casimir eigenvalues of the irreducible
representation. Such relations among diverse one-time systems provide
evidence that there is an underlying unifying principle behind them. The
principle is the {\it local} Sp$\left( 2,R\right) $ symmetry, and its
unavoidable consequence of demanding a spacetime with two timelike
dimensions which provides a basis for the {\it global} symmetry.

To describe spinning systems, worldline fermions $\psi _{\alpha }^{M}\left(
\tau \right) $, with $\alpha =1,2,\cdots ,n$ are introduced. Together with $%
X^{M},P^{M}$, they form the fundamental representation $\left( \psi _{\alpha
}^{M},X^{M},P^{M}\right) $ of the supergroup OSp$\left( n/2\right) $.
Gauging this supergroup \cite{spin} instead of Sp$\left( 2,R\right) $
produces a Lagrangian that has $n$ local supersymmetries plus $n$ local
conformal supersymmetries on the worldline, in addition to local Sp$\left(
2,R\right) $ and local SO$\left( n\right) $. The full set of first class
constraints that correspond to the generators of these gauge
(super)symmetries are, at the classical level, 
\begin{equation}
X\cdot X=P\cdot P=X\cdot P=X\cdot \psi _{\alpha }=P\cdot \psi _{\alpha
}=\psi _{\lbrack \alpha }\cdot \psi _{\beta ]}=0.  \label{ospn2gen}
\end{equation}
The classical solution of these constraints, with a flat spacetime metric $%
\eta ^{MN},$ require a signature with two timelike dimensions. Therefore, as
in the spinless case the global symmetry of the theory is SO$\left(
d,2\right) $. It is applied to the label $M$ on $\left( \psi _{\alpha
}^{M},X^{M},P^{M}\right) $. The global SO$\left( d,2\right) $ generators $%
J^{MN}$ that commute with all the OSp$\left( n/2\right) $ gauge generators (%
\ref{ospn2gen}) now include the spin 
\begin{equation}
J^{MN}=L^{MN}+S^{MN}\,,\quad S^{MN}=\frac{1}{2i}\left( \psi _{\alpha
}^{M}\psi _{\alpha }^{N}-\psi _{\alpha }^{N}\psi _{\alpha }^{M}\right) .
\label{SIJ}
\end{equation}
As in the spinless case, by gauge fixing the bosons as well as the fermions,
one finds a multitude of spinning one-time dynamical systems that are
unified by the same two-time system both at the classical and quantum
levels. All of these have SO$\left( d,2\right) $ hidden symmetry realized in
the same representation, where the representation is different for each $n$
(number of local supersymmetries on the worldline, which is related also to
the spin of the particle).

\section{Interactions with background fields}

The simple action in (\ref{confor}) is written in a flat two-time spacetime
with metric $\eta _{MN}$ which could be characterized as a ``free'' theory.
Interactions in the one-time systems emerged because of the first class
constraints $X^{2}=P^{2}=X\cdot P=0$, not because of explicit interactions
in the two time theory. The constraints generate the Sp$\left( 2,R\right) $
gauge symmetry. This symmetry was realized linearly on the doublet $%
X_{i}^{M}=\left( X^{M},P^{M}\right) $ and its generators were $%
Q_{ij}^{0}=X_{i}\cdot X_{j}.$

We now generalize the ``free'' theory to an ``interacting'' theory by
including background gravitational and gauge fields and other potentials.
This will be done by generalizing the worldline Hamiltonian (canonical
conjugate to $\tau $) $Q_{22}^{0}=P_{M}P_{N}\,\eta ^{MN}$ to a more general
form that includes a metric $G^{MN}\left( X\right) ,$ a gauge potential%
\footnote{%
It is possible to generalize this discussion by promoting $A$ to a
non-Abelian Yang-Mills potential coupled to a non-Abelian charge, which is
an additional dynamical degree of freedom. To keep the discussion simple we
take an Abelian $A$ in the present paper.} to gauge-covariantize the
momentum $P_{M}+A_{M}\left( X\right) ,$ and an additional potential $U\left(
X\right) $ that is added to the kinetic term. Generalizing $Q_{22}$ in this
way requires also generalizing all $Q_{ij}^{0}$ to $Q_{ij}\left( X,P\right) $
whose functional form will be determined. The Lagrangian is formally similar
to the ``free'' case (\ref{confor}) 
\begin{equation}
S=\int d\tau (\partial _{\tau }X^{M}P_{M}-\frac{1}{2}A^{ij}Q_{ij}\left(
X,P\right) \,).
\end{equation}
Whatever the expressions for $Q_{ij}\left( X,P\right) $ are, by the
equations of motion of the gauge potentials $A^{ij},$ they are required to
form first class constraints that close under the Sp$\left( 2,R\right) $
commutation rules (\ref{sp2alg}), which should follow from the basic
commutation rules of $\left( X^{M},P^{M}\right) .$ Furthermore, the local Sp$%
\left( 2,R\right) $ transformation properties of the dynamical variables
should be given by these generators under commutation rules 
\begin{eqnarray}
\delta X^{M} &=&\frac{i}{2}\omega ^{ij}\left( \tau \right) \left[
Q_{ij}\left( X,P\right) ,X^{M}\right] =\frac{1}{2}\omega ^{ij}\left( \tau
\right) \frac{\partial Q_{ij}\left( X,P\right) }{\partial P_{M}}  \label{t1}
\\
\delta P^{M} &=&i\omega ^{ij}\left( \tau \right) \left[ Q_{ij}\left(
X,P\right) ,P^{M}\right] =-\frac{1}{2}\omega ^{ij}\left( \tau \right) \frac{%
\partial Q_{ij}\left( X,P\right) }{\partial X^{M}} \\
\delta A^{ij} &=&\partial _{\tau }\omega ^{ij}+\omega ^{ik}\varepsilon
_{kl}A^{lj}+\omega ^{jk}\varepsilon _{kl}A^{li}.  \label{t2}
\end{eqnarray}
These certainly hold for the free case with $Q_{ij}^{0}=X_{i}\cdot X_{j},$
but now we discuss the general case. Substituting these transformation laws
into the Lagrangian we have (ignoring orders of operators at the classical
level) 
\begin{equation}
\delta L=\partial _{\tau }\left( \delta X^{M}\right) P_{M}+\partial _{\tau
}X^{M}\delta P_{M}-\frac{1}{2}\delta A^{ij}Q_{ij}\left( X,P\right) -\frac{1}{%
2}A^{ij}\delta Q_{ij}\left( X,P\right) \,
\end{equation}
where $\delta Q_{ij}\left( X,P\right) =\frac{\partial Q_{ij}}{\partial X^{M}}%
\delta X^{M}+\frac{\partial Q_{ij}}{\partial P_{M}}\delta P_{M}$. After an
integration by parts of the first term, using (\ref{t1}-\ref{t2}) this
becomes 
\begin{equation}
\delta L=-\frac{1}{2}\partial _{\tau }\left( \omega ^{ij}Q_{ij}\right) -%
\frac{1}{2}\left( \omega ^{ik}\varepsilon _{kl}A^{lj}+\omega
^{jk}\varepsilon _{kl}A^{li}\right) Q_{ij}-\frac{1}{4}A^{ij}\omega
^{kl}\left\{ Q_{ij},Q_{kl}\right\} ,
\end{equation}
where $\left\{ Q_{ij},Q_{kl}\right\} $ is the Poisson bracket 
\begin{equation}
\left\{ Q_{ij},Q_{kl}\right\} =\frac{\partial Q_{ij}}{\partial X^{M}}\frac{%
\partial Q_{kl}}{\partial P_{M}}-\frac{\partial Q_{ij}}{\partial P_{M}}\frac{%
\partial Q_{kl}}{\partial X^{M}}.
\end{equation}
Thus, if the $Q_{ij}$ satisfy the Sp$\left( 2,R\right) $ algebra (\ref
{sp2alg}), then the Poisson bracket term cancels the second term, and $%
\delta L$ is a total derivative. Hence to insure the gauge invariance of the
action $S$ we must require the differential constraints 
\begin{equation}
\frac{\partial Q_{ij}}{\partial X^{M}}\frac{\partial Q_{kl}}{\partial P_{M}}-%
\frac{\partial Q_{ij}}{\partial P_{M}}\frac{\partial Q_{kl}}{\partial X^{M}}%
=\varepsilon _{jk}Q_{il}+\varepsilon _{ik}Q_{jl}+\varepsilon
_{jl}Q_{ik}+\varepsilon _{il}Q_{jk}.  \label{poisson}
\end{equation}
With these restrictions we look for $Q_{ij}\left( X,P\right) $ that can be
interpreted as dynamics with background fields, as opposed to dynamics in
flat spacetime. To be able to integrate out the momenta $P^{M}$ we restrict
these expressions to contain at the most two powers of $P^{M}$ (this
restriction could be lifted to construct even more general systems \footnote{%
The coefficients of higher powers of $P^M$ have the interpretation of higher
spin fields}). Also, keeping the analogy to the flat case, we will take $%
Q_{11}$ to have no powers of $P^{M},$ $Q_{12}$ to have at most one power of $%
P^{M},$ and $Q_{22} $ to have at the most two powers of $P^{M},$ as follows 
\begin{eqnarray}
Q_{11} &=&W\left( X\right) ,\quad Q_{12}=\frac{1}{2}V^{M}\left(
P_{M}+A_{M}\right) +\frac{1}{2\sqrt{G}}\left( P_{M}+A_{M}\right) \sqrt{G}%
V^{M},\quad  \label{q1} \\
Q_{22} &=&\frac{1}{\sqrt{G}}\left( P_{M}+A_{M}\right) \sqrt{G}G^{MN}\left(
P_{N}+A_{N}\right) +U\left( X\right) .  \label{q2}
\end{eqnarray}
The functions $W\left( X\right) ,V^{M}\left( X\right) ,G^{MN}\left( X\right)
,A_{N}\left( X\right) ,U\left( X\right) $ will satisfy certain constraints.
The expression for $Q_{22}$ is a generalization of the free worldline
``Hamiltonian'' in flat space $\eta ^{MN}P_{M}P_{N}.$ The factors of $\sqrt{G%
}$ are inserted to insure hermiticity of the operators in a quantum theory
as applied on wavefunctions with a norm $\int \sqrt{G}\psi ^{\ast }\psi $.
In the classical theory the factors of $\sqrt{G}$ in $Q_{12},Q_{22}$ cancel
since orders of operators are neglected, but in any case a reordering
amounts to a redefinition of $A_{M}\left( X\right) $ and $U\left( X\right) $.

The combination $P_{M}+A_{M}\left( X\right) $ is gauge invariant under $%
\delta _{\Lambda }A_{M}\left( X\right) =\partial _{M}\Lambda \left( X\right) 
$ and $\delta _{\Lambda }P_{M}=-\partial _{M}\Lambda \left( X\right) $,
where $\Lambda \left( X\left( \tau \right) \right) $ is a gauge function of
spacetime. The Lagrangian has this gauge symmetry since it transforms into a
total derivative under the gauge transformation $\delta _{\Lambda
}L=-\partial _{\tau }X^{M}\partial _{M}\Lambda \left( X\right) =-\partial
_{\tau }\Lambda $. Furthermore, the Lagrangian is a scalar under spacetime
general coordinate transformations, since the $Q_{ij}$ are scalars when all
the background fields are transformed as tensors, while the term $\partial
_{\tau }X^{M}P_{M}$ is invariant under $\delta _{\varepsilon
}X^{M}=-\varepsilon ^{M}\left( X\right) $ and $\delta _{\varepsilon
}P_{M}=\partial _{M}\varepsilon ^{N}P_{N}$. Of course, if the background
fields are fixed, the general covariance and gauge symmetries are not
generally valid, and only a subgroup that corresponds to Killing symmetries
of the combined gauge and reparametrization transformations survive.

By integrating out $P_{M}$ we can rewrite the Lagrangian purely in terms of $%
X^{M}\left( \tau \right) $ and its derivatives $\dot{X}^{M}\left( \tau
\right) $ 
\begin{equation}
L=\frac{1}{2A^{22}}\left( \dot{X}^{M}-A^{12}V^{M}\right) G_{MN}\,\left( \dot{%
X}^{N}-A^{12}V^{N}\right) -\frac{A^{22}}{2}U-\frac{A^{11}}{2}W-\dot{X}%
^{M}A_{M}.  \label{actx}
\end{equation}
By inspection of (\ref{q2}) or (\ref{actx}) we interpret $A_{M}\left(
X\right) $ as a gauge field, $G_{MN}\left( X\right) $ as a spacetime metric
and $U\left( X\right) $ as an additional potential. The function $W\left(
X\right) \sim 0$ is the constraint that replaces $X\cdot X\sim 0$ and the
vector $V^{M}\left( X\right) $ can be thought of as a general coordinate
transformation since the action of $Q_{12}$ on phase space is $\delta
_{12}X^{M}=V^{M}\left( X\right) $ and $\delta _{12}P_{M}=\partial
_{M}V^{K}P_{K}+\partial _{M}\left( V\cdot A\right) $ which looks like a
general coordinate transformation up to a gauge transformation$.$

The classical local Sp$\left( 2,R\right) $ transformation laws for $\left(
X^{M},P_{M}\right) $ in phase space follow from $\left( \ref{t1},\ref{t2}%
\right) $ 
\begin{eqnarray}
\delta X^{M} &=&\omega ^{12}\left( \tau \right) V^{M}+\omega ^{22}\left(
\tau \right) G^{MN}\left( P_{N}+A_{N}\right)  \\
\delta P_{M} &=&-\frac{1}{2}\omega ^{11}\left( \tau \right) \partial
_{M}W-\omega ^{12}\left( \tau \right) \left[ \left( \partial
_{M}V^{N}\right) P_{N}+\partial _{M}\left( V\cdot A\right) \right]  \\
&&-\frac{1}{2}\omega ^{22}\left( \tau \right) \left[ \left( \partial
_{M}G^{KL}\right) \left( P_{K}+A_{K}\right) \left( P_{L}+A_{L}\right)
+\partial _{M}U+2G^{KL}\partial _{M}A_{K}\left( P_{L}+A_{L}\right) \right]  
\nonumber
\end{eqnarray}
This, together with (\ref{t2}), is a local symmetry of the action provided (%
\ref{poisson}) is satisfied. These conditions give the following
differential constraints on the functions $W\left( X\right) ,$ $V^{M}\left(
X\right) ,$ $G^{MN}\left( X\right) ,$ $A_{N}\left( X\right) ,$ $U\left(
X\right) $. From $\left\{ Q_{11},Q_{22}\right\} =4Q_{12}$ we learn 
\begin{equation}
V^{M}=\frac{1}{2}G^{MN}\partial _{N}W.  \label{one}
\end{equation}
From $\left\{ Q_{11},Q_{12}\right\} =2Q_{11}$ we learn 
\begin{equation}
V^{M}\partial _{M}W=2W,\quad or\quad G^{MN}\left( \partial _{M}W\right)
\left( \partial _{N}W\right) =4W.  \label{two}
\end{equation}
Finally from $\left\{ Q_{22},Q_{12}\right\} =-2Q_{22}$ we learn (from the
coefficients of each power of $P_{M})$ that 
\begin{equation}
\pounds _{V}G^{MN}=-2G^{MN},\quad V^{M}\partial _{M}U=-2U,\quad
V^{M}F_{MN}=0,  \label{homo}
\end{equation}
where $\pounds _{V}G^{MN}$ is the Lie derivative of $G^{MN}$ (an
infinitesimal general coordinate transformation) 
\begin{equation}
\pounds _{V}G^{MN}\equiv V^{K}\partial _{K}G^{MN}-\partial
_{K}V^{M}G^{KN}-\partial _{K}V^{N}G^{MK},
\end{equation}
and $F_{MN}=\partial _{M}A_{N}-\partial _{N}A_{M}$ is the gauge field
strength. The differential equation $\pounds _{V}G^{MN}=-2G^{MN}$ \ together
with (\ref{one}) was called a ``closed homothety'' condition on the geometry 
\footnote{%
I learned this term when I came across ref.\cite{strominger}, after having
derived these equations independently sometime ago. The physical problem in
the present paper is quite different than \cite{strominger} where our
spacetime index $M$ (with (d,2) signature) is replaced by a particle label
for multiparticles in \cite{strominger} (with Euclidean signature);
nevertheless the mathematics formally coincide with ref.\cite{strominger}.
After the current paper was submitted for publication, I was informed that
similar equations were obtained in \cite{sezgintanii} in the context of
conformally invariant sigma models on a p+1 dimensional worldvolume, using a
very different approach than ours. Although the case of p=0 (worldline)
relevant for our case was missed by these authors, when their expressions
are continued to p=0 they agree with our results. While there are formal
similarities, an important difference between our work and those of \cite
{strominger} and \cite{sezgintanii} is that we have local SO(1,2)=Sp(2)
symmetry as opposed to their global symmetry. This requires the constraints $%
Q_{ij}\left( X,P\right) =0$ which demand a spacetime with two timelike
dimensions, thus leading to conceptually very different physics.}. We have
the added generalization of the gauge field $A_{M}$ in our case. When all
fields are present they are coupled to each other.

The differential equation for the gauge field may also be rewritten in terms
of the Lie derivative on the vector $\pounds _{V}A_{M}=\partial _{M}\left(
V\cdot A\right) $, where the Lie derivative on the vector is $\pounds
_{V}A_{M}=V^{K}\partial _{K}A_{M}+\partial _{M}V^{K}A_{K}$ (an infinitesimal
general coordinate transformation). Using the gauge invariance of the
physics, without loss of generality one may choose an axial gauge $V\cdot
A=0 $. There still is a remaining gauge symmetry $\delta _{\Lambda
}A_{M}=\partial _{M}\Lambda $, for all $\Lambda $ that satisfy $%
V^{K}\partial _{K}\Lambda =0$. Thus, the gauge field equation may be
rewritten in the form 
\begin{equation}
\pounds _{V}A_{M}=0,\quad V\cdot A=0,  \label{gauge}
\end{equation}
with a remaining gauge symmetry of these equations $\left\{ \Lambda
;\,\,V^{K}\partial _{K}\Lambda =0\right\} $ which we will make use of later.

Any solution to the coupled equations (\ref{one}, \ref{two}, \ref{homo}, \ref
{gauge}) gives an action with local Sp$\left( 2,R\right) $ symmetry. Such an
action provides a two-time physics theory including interactions with
background fields. The global symmetries correspond to Killing symmetries in
the presence of backgrounds, which is a subgroup embedded in general
coordinate transformations combined with gauge transformations. This is the
global symmetry, which in the flat and free case becomes SO$\left(
d,2\right) $.

The Sp$\left( 2,R\right) $ gauge symmetry may be gauge fixed to define a
``time'' and analyze the system from the point of view of one-time physics.
The global symmetry described in the previous paragraph survives after gauge
fixing the Sp$\left( 2,R\right) $ local symmetry, since it commutes with it
(recall the $Q_{ij}$ are invariant under general coordinate and gauge
transformations). This global symmetry would then become the non-linearly
realized hidden global symmetries in each of the one-time dynamical systems
that emerge after gauge fixing (in the ``free'' case it is SO$\left(
d,2\right) $). The symmetry must be realized in the same representation for
each one-time dynamical system that belongs to the same class, where the
class is fixed by a given set of background fields.

\section{Pure gauge field background}

When the background metric is flat $G^{MN}=\eta ^{MN}$ the only solution of
the homothethy condition $\pounds _{V}G^{MN}=-2G^{MN}$ \ is $V^{M}=X^{M}.$
This immediately gives $W=X\cdot X,$ and $U$ is any homogeneous function of $%
X^{M}$ of degree -2. The global symmetry of the metric is SO$\left(
d,2\right) .$ If we want to keep the SO$\left( d,2\right) $ symmetry, $U$
could only be $U=g/X\cdot X$ (however, without the SO$\left( d,2\right) $
symmetry one can allow some other $U$ of degree -2).

The equations for the gauge field (\ref{gauge}) simplify in flat space. The
remaining gauge symmetry parameter is homogeneous of degree zero $X\cdot
\partial \Lambda =0$ in $d+2$ dimensions. This is sufficient to fix further
the gauge $\partial _{M}A^{M}=0$ since according to the equations $A_{M}$
also is homogeneous of degree $-1$ in this gauge. The three equations
satisfied by the gauge field are now 
\begin{equation}
X\cdot A\left( X\right) =0,\quad \left( X\cdot \partial +1\right)
A_{M}\left( X\right) =0,\quad \partial _{M}A^{M}=0.  \label{diraceq}
\end{equation}
There still remains gauge symmetry in these equations for $\Lambda $ that
satisfy $X\cdot \partial \Lambda =\partial \cdot \partial \Lambda =0.$ The
content of these equations for $\Lambda $ is still non-trivial.

These equations were proposed by Dirac in 1936 \cite{Dirac} as subsidiary
conditions to describe the usual 4-dimensional Maxwell theory of
electromagnetism (in the Lorentz gauge), as a theory in 6 dimensions which
automatically displays SO$\left( 4,2\right) $ symmetry. Dirac's aim was to
linearize the conformal symmetry of the 4 dimensional Maxwell theory. The
subsidiary conditions can be regarded as ``kinematics'' while dynamics is
given by a Klein-Gordon type equation in 6-dimensions that may include
interactions with other fields. As Dirac showed, the linear SO$\left(
4,2\right) $ Lorentz symmetry of the 6 dimensional theory is indeed the
non-linear conformal symmetry of the Maxwell theory.

Actually, in the framework of two-time physics, conformal symmetry is only
one of the possible interpretations of the SO$\left( 4,2\right) $ global
symmetry of these equations. In two-time physics this interpretation relies
on a particular choice of ``time'' among the two available timelike
dimensions, while with other gauge choices the interpretation of the SO$%
\left( 4,2\right) $ symmetry is completely different than conformal
symmetry. To illustrate this, denote the components of the 6 dimensions as $%
X^{M}=\left( X^{+^{\prime }},X^{-^{\prime }},X^{\mu }\right) $ with metric $%
X\cdot X=-2X^{+^{\prime }}X^{-^{\prime }}+X_{\mu }X^{\mu }.$ The Sp$\left(
2,R\right) $ gauge choices $P^{+^{\prime }}\left( \tau \right) =0$, $%
X^{+^{\prime }}\left( \tau \right) =1$ eliminates one timelike and one
spacelike dimensions and brings down the two-time formulation in $d+2$
dimension to a one time formulation in $d$ dimensions. It is convenient to
use the electromagnetic gauge choice $A^{+^{\prime }}\left( X\right) =0$
(instead of Dirac's $\partial _{M}A^{M}=0$). Then the solution of the gauge
choices and constraints (including $Q_{11}=Q_{12}=0$), $X\cdot X=X\cdot
P=X\cdot A=0,$ is given in the following form 
\begin{eqnarray}
X^{M}\left( \tau \right) &=&\left( 1,\,\,x^{2}/2,\,\,x^{\mu }\left( \tau
\right) \right) ,\quad P^{M}=\left( 0,\,x\cdot p,\,p^{\mu }\left( \tau
\right) \right) , \\
A^{M}\left( X\right) &=&\left( \,0,\,\,x\cdot A,\,\,A^{\mu }\left( x\left(
\tau \right) \right) \,\,\right) .
\end{eqnarray}
The dynamics of the remaining degrees of freedom $\left( x^{\mu }\left( \tau
\right) ,p^{\mu }\left( \tau \right) \right) $ are obtained by substituting
these solutions into the gauge invariant 6-dimensional action (\ref{actx}).
The result is the standard 4-dimensional action for the massless
relativistic particle coupled to the electromagnetic gauge potential $A_{\mu
}\left( x\right) $ 
\begin{equation}
L=\frac{1}{2A^{22}}\left( \dot{x}^{\mu }\right) ^{2}-\dot{x}^{\mu }A_{\mu
}\left( x\right) .
\end{equation}
Thus the original two-time action displays explicitly the hidden SO$\left(
4,2\right) $ symmetry of the one-time action. The general coordinate
transformation of the previous section, specialized to $\varepsilon
^{M}=\varepsilon ^{MN}X_{N}$ with constant antisymmetric $\varepsilon ^{MN},$
is the SO$\left( 4,2\right) $ global Lorentz symmetry of the 6-dimensional
action, including the gauge field. This 6-dimensional Lorentz symmetry is
also the non-linearly realized conformal symmetry of the gauge fixed action
above, since the global symmetry commutes with the gauge symmetry, and gauge
fixing of the gauge invariant action could not destroy the global symmetry.
Indeed the generators of conformal transformations are the gauge invariant $%
L^{MN}=X^{M}P^{N}-X^{N}P^{N}$ now expressed in terms of the gauge fixed
coordinates and momenta as shown in \cite{conf}\cite{lifting}. This agrees
with Dirac's interpretation of the conformal SO$\left( 4,2\right) $ symmetry
as being the Lorentz symmetry of 6 dimensions.

However, if one chooses another gauge for time instead of $X^{+^{\prime
}}\left( \tau \right) =1$, as was done with many illustrations in \cite{conf}
\cite{lifting}, other $d$-dimensional dynamical systems arise, which now are
coupled to a gauge potential. Then the SO$\left( d,2\right) $ symmetry
generated by the same $L^{MN}$ has a different interpretation than conformal
symmetry, as explained in \cite{conf}\cite{lifting}. The presence of the
gauge field background now produces a large class of dynamical systems with
hidden SO$\left( d,2\right) $ symmetries, and Sp$\left( 2,R\right) $ duality
relations among them.

The two-time physics approach \cite{conf}-\cite{string2t} was developed
without being aware of the field equations invented by Dirac. While Dirac
was interested in linearizing conformal symmetry\footnote{%
I thank Vasilev for informing me of Dirac's work and the line of research
that followed the same trend of thought in relation to conformal symmetry 
\cite{Dirac}\cite{Kastrup}\cite{salam}\cite{vasilev}. A field theoretic
formulation of two-time physics has been derived recently \cite{fieldth} and
its relation to Dirac's work has been established. It is shown in \cite
{fieldth} that two-time physics in a field theoretic setting, as in the
particle dynamics setting, unifies different looking one-time field theories
as being the same two-time field theory, while simultaneously revealing
previously unnoticed hidden symmetries in field theory, including
interactions. Such duality and global symmetry properties of two-time
physics go well beyond Dirac's goal of linearizing conformal symmetry.}, the
motivation for the work in \cite{conf}-\cite{string2t} came independently
from duality, and signals for two-timelike dimensions in M-theory and its
extended superalgebra including D-branes \cite{ibtokyo}\cite{vafa}\cite
{stheory}. Driven by different motivations, and unaware of Dirac's approach
to conformal symmetry, two-time physics produced new insights that include
conformal symmetry but go well beyond it. Besides providing a deeper Sp$%
\left( 2,R\right) $ gauge symmetry as the fundamental basis for Dirac's
approach (see further \cite{fieldth}), two-time physics unifies classes of
one-time physical systems in $d$ dimensions that previously would have been
thought of as being unrelated to each other. The SO$\left( d,2\right) $
symmetry is interpreted as conformal symmetry in a certain one-time system,
but in other dually related dynamical systems it is a hidden symmetry with a
different interpretation, but realized in exactly the same irreducible
representation. . The unifying aspect in all the interpretations is that the
symmetry is the underlying spacetime symmetry in a spacetime that includes
two timelike dimensions.

\section{Gravitational background}

We now seek a solution of (\ref{one}-\ref{gauge}) that includes gravity in $%
d $ dimensions. It is convenient to make a change of variables $%
X^{M}=X^{M}\left( \kappa ,w,x^{\mu }\right) $ such that the function $%
W\left( X\right) $ is identified with the product of new coordinates $%
-2w\kappa $, while the coordinate $x^{\mu }$ is in $d$ dimensions. The
inverse of this change of variables is, $\kappa =K\left( X\right) ,$ $%
w=-W\left( X\right) /2K\left( X\right) $ and $x^{\mu }=x^{\mu }\left(
X\right) .$ Before we look for a solution to (\ref{one}-\ref{gauge}) it is
instructive to consider the example of the flat case that has components $%
X^{M}=\left( X^{+^{\prime }},X^{-^{\prime }},X^{\mu }\right) $ with the
constraint $W\left( X\right) =X\cdot X=-2X^{+^{\prime }}X^{-^{\prime
}}+X_{\mu }X^{\mu }.$ The change of variables and the inverse relations for
this case are 
\begin{eqnarray}
X^{+^{\prime }} &=&\kappa ,\quad X^{-^{\prime }}=\frac{\kappa x^{2}}{2}%
+w,\quad X^{\mu }=\kappa x^{\mu } \\
\kappa &=&X^{+^{\prime }},\quad w=\frac{X\cdot X}{-2X^{+^{\prime }}},\quad
x^{\mu }=\frac{X^{\mu }}{X^{+^{\prime }}}
\end{eqnarray}
This change of variables is a special case of a general coordinate
transformation. The flat metric in the new variables takes the form 
\begin{eqnarray}
ds^{2} &=&dX^{M}dX^{N}\eta _{MN}=-2dX^{+^{\prime }}dX^{-^{\prime }}+dX^{\mu
}dX^{\nu }\eta _{\mu \nu } \\
&=&-2d\kappa dw+\kappa ^{2}dx^{\mu }dx^{\nu }\eta _{\mu \nu }.
\end{eqnarray}
For this choice of basis we have $V^{M}=\left( \kappa ,w,0\right) $ and $%
W=-2\kappa w$ and the homothety conditions are easily verified. Taking this
form as a model we seek a similar solution. With a choice of coordinates we
can always take $V^{M}=\left( \kappa ,w,0\right) $. In the new coordinate
system $W\left( \kappa ,w,x^{\mu }\right) $ needs to be determined
consistently with the closed homothety conditions. We will make an ansatz
which may not be the most general, but is adequate to provide a sufficiently
large set of solutions. Thus, we will take $W\left( \kappa ,w,x\right)
=-2w\kappa $ to have the same form as the free case, and insert these forms
of $V,W$ in the closed homothety conditions with a general $G^{MN}$. The
homothety condition reads 
\begin{equation}
\left( \kappa \partial _{\kappa }+w\partial _{w}\right) G^{MN}-\delta
_{\kappa }^{M}G^{\kappa N}-\delta _{w}^{M}G^{wN}-\delta _{\kappa
}^{N}G^{\kappa M}-\delta _{w}^{N}G^{wM}=-2G^{MN}.
\end{equation}
From $V^{M}=\frac{1}{2}G^{MN}\partial _{N}W\left( X\right) $ we learn
further 
\begin{eqnarray}
V^{\mu } &=&0=-G^{\mu \kappa }w-G^{\nu w}\kappa \quad \rightarrow G^{\mu
\kappa }=\frac{1}{\kappa }W^{\mu },\quad G^{\mu w}=-\frac{w}{\kappa ^{2}}%
W^{\mu }, \\
V^{\kappa } &=&\kappa =-G^{\kappa \kappa }w-G^{\kappa w}\kappa ,\quad
\rightarrow \ G^{\kappa \kappa }=-\frac{\kappa }{w}\left( 1+G^{\kappa
w}\right) \\
V^{w} &=&w=-G^{w\kappa }w-G^{ww}\kappa ,\quad \rightarrow G^{ww}=-\frac{w}{%
\kappa }\left( 1+G^{\kappa w}\right)
\end{eqnarray}
Specializing the indices in the homothety condition gives the solutions for
all components of $G^{MN}$ in the form 
\begin{equation}
G^{MN}=\left( 
\begin{array}{ccc}
\frac{\kappa }{w}\left( \gamma -1\right) & -\gamma & \frac{1}{\kappa }W^{\nu
} \\ 
-\gamma & \frac{w}{\kappa }\left( \gamma -1\right) & -\frac{w}{\kappa ^{2}}%
W^{\nu } \\ 
\frac{1}{\kappa }W^{\mu } & -\frac{w}{\kappa ^{2}}W^{\mu } & \frac{g^{\mu
\nu }}{\kappa ^{2}}
\end{array}
\right)
\end{equation}
where the functions $\gamma \left( x,\frac{w}{\kappa }\right) ,$ $W^{\mu
}\left( x,\frac{w}{\kappa }\right) ,$ $g^{\mu \nu }\left( x,\frac{w}{\kappa }%
\right) $ are arbitrary functions of only $x^{\mu }$ and the ratio $\frac{w}{%
\kappa }$.

In this coordinate system we can also solve the kinematic conditions for the
gauge field (\ref{gauge}), which become 
\begin{equation}
\left( w\partial _{w}+\kappa \partial _{\kappa }\right) A_{M}+\delta
_{M}^{w}A_{w}+\delta _{M}^{\kappa }A_{\kappa }=0,\quad wA_{w}+\kappa
A_{\kappa }=0.
\end{equation}
The general solution is 
\[
A_{w}=\frac{1}{\kappa }B\left( \frac{w}{\kappa },x\right) ,\quad A_{\kappa
}=-\frac{w}{\kappa \,^{2}}B\left( \frac{w}{\kappa },x\right) ,\quad A_{\mu
}=A_{\mu }\left( \frac{w}{\kappa },x\right) . 
\]
The remaining gauge symmetry $V^{M}\partial _{M}\Lambda =0$ is just
sufficient to set $B=0$ in this solution, if so desired. Finally the
solution for $U\left( w,\kappa ,x\right) $ that satisfies $V^{M}\partial
_{M}U=-2U$ is 
\begin{equation}
U=\frac{1}{\kappa ^{2}}u\left( \frac{w}{\kappa },x\right) .
\end{equation}

For this solution, the generators of Sp$\left( 2,R\right) $ in $\left( \ref
{q1},\ref{q2}\right) $ become, in the gauge $B=0$, 
\begin{eqnarray}
Q_{11} &=&-2\kappa w,\quad Q_{12}=\kappa p_{\kappa }+wp_{w}, \\
Q_{22} &=&-2\gamma p_{w}p_{\kappa }+\left( p_{\kappa }^{2}\frac{\kappa }{w}%
+p_{w}^{2}\frac{w}{\kappa }\right) \left( \gamma -1\right) \\
&&+\frac{2}{\kappa ^{2}}\left( \kappa p_{\kappa }-wp_{w}\right) W^{\mu
}p_{\mu }+\frac{H}{\kappa ^{2}},  \nonumber
\end{eqnarray}
where 
\begin{equation}
H=\frac{1}{\sqrt{-g}}\left( p_{\mu }+A_{\mu }\right) \sqrt{-g}g^{\mu \nu
}\left( p_{\nu }+A_{\nu }\right) +u.
\end{equation}
It is easy to verify directly that they close correctly for any background
fields $\gamma ,g_{\mu \nu },W^{\mu },A_{\mu },u$ that are arbitrary
functions of $\left( \frac{w}{\kappa },x^{\mu }\right) $.

Imposing the Sp$\left( 2,R\right) $ constraints $Q_{ij}=0$ is now easy. It
is convenient to choose a Sp$\left( 2,R\right) $ gauge, which we know will
produce a one-time theory. A gauge choice that is closely related to the
massless relativistic particle is taken by analogy to the flat theory. At
the classical level we choose the Sp$\left( 2,R\right) $ gauges $\kappa
\left( \tau \right) =1$ and $p_{w}\left( \tau \right) =0,$ and solve $%
Q_{11}=Q_{12}=0$ in the form $w\left( \tau \right) =p_{\kappa }\left( \tau
\right) =0$. There remains unfixed one gauge subgroup of Sp$\left(
2,R\right) $ which corresponds to $\tau $ reparametrization, and the
corresponding Hamiltonian constraint $H\sim 0$, which involves the
background fields $g_{\mu \nu }\left( x\right) ,$ $A_{\mu }\left( x\right) ,$
$u\left( x\right) $ that now are functions of only the $d$ dimensional
coordinates $x^{\mu },$ since $w/\kappa =0.$ In this gauge, the background
fields $\gamma ,W^{\mu }$ decouple from the dynamics that govern the time
development of $x^{\mu }\left( \tau \right) .$ The two-time theory described
by the original Lagrangian (\ref{actx}) reduces to a one-time theory 
\[
L=\frac{1}{2A^{22}}\dot{x}^{\mu }\dot{x}^{\nu }g_{\mu \nu }\left( x\right) -%
\frac{A^{22}}{2}u\left( x\right) -\dot{x}^{\mu }A_{\mu }\left( x\right) . 
\]
which controls the dynamics of the remaining degrees of freedom $x^{\mu
}\left( \tau \right) .$ Evidently this Lagrangian describes a particle
moving in arbitrary gravitational, electromagnetic gauge fields and other
potential $g_{\mu \nu }\left( x\right) ,$ $A_{\mu }\left( x\right) $, $%
u\left( x\right) $ in the remaining $d$ dimensional spacetime.

We have therefore demonstrated that {\it all usual interactions} experienced
by a particle, as described in the one-time formulation of dynamics, can be
embedded in two time physics as a natural solution of the two-time equations
(\ref{one}-\ref{gauge}), taken in a fixed Sp$\left( 2,R\right) $ gauge.

\section{Spinning particles in background fields}

To describe spinning particles in two time physics we need local
superconformal symmetry instead of local conformal symmetry, as demonstrated
in flat space in \cite{spin}. There the Sp$\left( 2,R\right) $ gauge group
was replaced by the supergroup OSp$\left( n|2\right) $ as described at the
end of section 2 of this paper. To generalize this approach to curved space
we need a soldering form $E_{M}^{a}$ and its inverse $E_{a}^{M}$ (analog of
vierbein) that transforms curved base space indices to flat tangent space
indices and vice versa. The metric in tangent space is $\eta _{ab}$ while
the general metric is given by $G_{MN}=E_{M}^{a}E_{N}^{b}\eta _{ab}.$ Next
consider phase space including spin degrees of freedom $\left(
X^{M},P_{M},\psi _{\alpha }^{a}\right) $ where $a$ is a tangent space index
and $\alpha =1,2,\cdots ,n$ denote the $n$ supersymmetries. The canonical
commutation rules are 
\begin{equation}
\left[ X^{M},P_{N}\right] =i\delta _{N}^{M},\quad \left\{ \psi _{\alpha
}^{a},\psi _{\beta }^{b}\right\} =\eta ^{ab}\delta _{\alpha \beta }.
\label{basiccom}
\end{equation}
The $\psi _{\alpha }^{a}$ form a Clifford algebra and may be represented by
gamma matrices if so desired.

A Lagrangian that has the desired OSp$\left( n|2\right) $ local symmetry has
the same form as the flat case given in \cite{spin} with some modifications 
\begin{equation}
L=\dot{X}^{M}P_{M}+\frac{i}{2}\psi _{\alpha }^{a}\dot{\psi}_{\alpha
}^{b}\eta _{ab}-\frac{1}{2}A^{ij}Q_{ij}+iF^{i\alpha }Q_{i\alpha }-\frac{1}{2}%
B^{\alpha \beta }Q_{\alpha \beta },
\end{equation}
The OSp$\left( n|2\right) $ gauge fields may be arranged into the form of a $%
\left( n+2\right) \times \left( n+2\right) $ supermatrix 
\begin{equation}
\left( 
\begin{array}{ll}
B^{\left[ \alpha \beta \right] } & F^{\alpha i} \\ 
\varepsilon _{ij}F^{j\beta } & A^{ij}
\end{array}
\right) ,\quad A,B=bose,\quad F=fermi
\end{equation}
They obey the standard transformation rules for gauge fields, as given in 
\cite{spin}. The OSp$\left( n|2\right) $ generators $Q_{ij},Q_{i\alpha
},Q_{\alpha \beta }$ are to be taken as non-linear functions in phase space,
including background fields. As in the purely bosonic case, our task is to
find the forms of the background fields that have an interpretation as
gravitational, gauge or other interactions experienced by spinning particles
in two-time physics. The gauge field equations of motion require the first
class constraints $Q_{ij}\sim Q_{i\alpha }\sim Q_{\alpha \beta }\sim 0,$
whose solution will require two timelike dimensions, as in the flat theory
or as in the curved purely bosonic theory. These are then the generators of
infinitesimal transformations that tell us how to transform $\delta
X^{M},\delta P_{M},\delta \psi _{\alpha }^{M}$ under the local OSp$\left(
n|2\right) .$ As in the purely bosonic theory treated earlier in this paper,
it is easy to show that the Lagrangian has the local symmetry provided these
first class constraints close into the algebra of OSp$\left( n|2\right) .$
This requirement gives the differential equations for the background fields.

In the flat case the OSp$\left( n|2\right) $ generators are given by $%
Q_{ij}^{0}=X_{i}\cdot X_{j},$ $Q_{i\alpha }^{0}=X_{i}\cdot \psi _{\alpha },$
and $Q_{\alpha \beta }^{0}=\frac{i}{2}\psi _{\lbrack \alpha }\cdot \psi
_{\beta ]}.$ To include background fields we first generalize the fermionic
generators $P\cdot \psi _{\alpha }$ ($n$ local supersymmetries) and $X\cdot
\psi _{\alpha }$ ($n$ local superconformal symmetries) by introducing a
tangent space vector $V_{a}\left( X\right) $, a soldering from $%
E_{M}^{a}\left( X\right) ,$ a spin connection $\omega _{M}^{ab}\left(
X\right) ,$ a gauge field $A_{M}\left( X\right) ,$ and replacing the
momentum by the covariant momentum 
\begin{equation}
\Pi _{a}\left( X,P,\psi \right) =E_{a}^{M}\left( P_{M}+A_{M}+\frac{1}{2}%
\omega _{M}^{ab}S_{ab}\right)
\end{equation}
The spin connection, which generally includes torsion, is coupled to the
spin operator $S^{ab}=\frac{1}{2i}\left( \psi _{\alpha }^{a}\psi _{\alpha
}^{b}-\psi _{\alpha }^{b}\psi _{\alpha }^{a}\right) $ to form the covariant
momentum. The generalized fermionic generators are as follows 
\begin{equation}
Q_{1\alpha }=\psi _{\alpha }^{a}V_{a}\left( X\right) ,\quad Q_{2\alpha }=%
\frac{1}{2}\left( \psi _{\alpha }^{a}\Pi _{a}+\tilde{\Pi}_{a}\psi _{\alpha
}^{a}\right) .
\end{equation}
The bosonic generators are computed from the closure of the OSp$\left(
n|2\right) $ commutation relations 
\begin{equation}
\left\{ Q_{1\alpha },Q_{1\beta }\right\} =\delta _{\alpha \beta
}Q_{11},\quad \left\{ Q_{2\alpha },Q_{2\beta }\right\} =\delta _{\alpha
\beta }Q_{22},\quad \left\{ Q_{1\alpha },Q_{2\beta }\right\} =\delta
_{\alpha \beta }Q_{12}+Q_{\alpha \beta }.  \label{closure}
\end{equation}
where $Q_{\alpha \beta }$ is the antisymmetric SO$\left( n\right) $
generator, and $Q_{ij}$ are the symmetric Sp$\left( 2\right) $ generators.
Note that $Q_{2\alpha }$ contains up to cubic terms in the fermions. $\tilde{%
\Pi}_{a}$ is given by $\tilde{\Pi}_{a}=\left( \sqrt{G}\right) ^{-1}\Pi _{a}%
\sqrt{G}$, where the factors of $\sqrt{G}$ insure hermiticity in a quantum
theory with correct factor ordering, but for the invariance of the classical
action, where we only need Poisson brackets instead of the commutators as
explained in the spinless case, these factors may be neglected.

For simplicity we will impose the flat $Q_{\alpha \beta }=Q_{\alpha \beta
}^{0}$ 
\begin{equation}
Q_{\alpha \beta }=\frac{i}{2}\psi _{\lbrack \alpha }\cdot \psi _{\beta ]}
\end{equation}
but will compute $Q_{ij}$ as a function of the background fields\footnote{%
We could have included also $E_{I}^{M}W_{M}^{ab}Q_{ab}^{0}$ as part of $\Pi
_{I},$ with $W_{M}^{ab}$ a gauge field that acts in the SO$\left( n\right) $
space within OSp$\left( n|2\right) .$ In that case we could also introduce a
vielbein $E_{a}^{A}$ for an internal space. For simplicity we will omit
these complications and seek a solution with a ``flat'' SO$\left( n\right) $
space, implying that the metric in SO$\left( n\right) $ space is $\delta
_{ab}$ instead of a curved space metric $G_{AB}=E_{A}^{a}E_{B}^{a}.$ Recall
that in the final analysis we are interested in imposing $Q_{ab}=0$ as part
of the singlet condition. I the presence of non-singlet background fields
such as $E_{A}^{a},W_{M}^{ab}$ this condition is harder to satisfy.}. This
condition requires that $E_{M}^{a}$ be determined in terms of $V^{a},\omega
_{M}^{ab}$%
\begin{equation}
E_{M}^{a}=D_{M}V^{a}=\partial _{M}V^{a}+\omega _{M}^{ab}V_{b},
\label{vielbein}
\end{equation}
while 
\begin{equation}
Q_{11}=V^{a}V^{b}\eta _{ab},\quad Q_{12}=\frac{1}{2}\left( V^{a}\Pi _{a}+%
\tilde{\Pi}_{a}V^{a}\right) ,\quad Q_{22}=\frac{1}{n}\left[ \frac{1}{2}%
\left( \psi _{\alpha }^{a}\Pi _{a}+\tilde{\Pi}_{a}\psi _{\alpha }^{a}\right) 
\right] ^{2}.
\end{equation}
Note that $Q_{22}$ contains several powers of the fermions. The closure (\ref
{closure}) is possible provided the gauge field strength and the curvature
are transverse to $V$ 
\begin{equation}
V^{M}F_{MN}=0,\quad V^{M}R_{MN}^{ab}=0,
\end{equation}
where 
\begin{equation}
V^{M}=E_{a}^{M}V^{a},  \label{v}
\end{equation}
and 
\begin{equation}
F_{MN}=\partial _{M}A_{N}-\partial _{N}A_{M}+\left[ A_{M},A_{N}\right]
,\quad R_{MN}^{ab}=\partial _{M}\omega _{N}^{ab}-\partial _{N}\omega
_{M}^{ab}+\left[ \omega _{M},\omega _{N}\right] ^{ab}.
\end{equation}
Furthermore, since $E_{M}^{a}=D_{M}V^{a}$ the torsion is determined in terms
of the curvature and $V$ as 
\begin{equation}
\quad T_{MN}^{a}=D_{M}E_{N}^{a}-D_{N}E_{M}^{a}=R_{MN}^{ab}V_{b},
\end{equation}
and is automatically transverse to $V$ provided the curvature is.

There remains to check the Sp$\left( 2\right) \times SO\left( n\right) $
closure of the bosonic generators. The SO$\left( n\right) $ part is trivial.
The Sp$\left( 2\right) $ part is similar to the purely bosonic case of the
previous section and is subject to the same conditions (\ref{one})-(\ref
{homo}) discussed there. However now $W,G^{MN}$ are given by $W=V^{a}V_{a}$
and $G_{MN}=E_{M}^{a}E_{N}^{b}\eta _{ab}$ and $U=0.$ These forms
automatically satisfy (\ref{one})-(\ref{homo}) provided $E_{M}^{a}$ is of
the form (\ref{vielbein}). In particular, (\ref{one}) is satisfied as
follows 
\begin{equation}
V^{M}=\frac{1}{2}G^{MN}\partial _{N}W=G^{MN}\left( D_{N}V^{a}\right)
V_{a}=G^{MN}E_{N}^{a}V_{a}=E_{b}^{M}V^{b}
\end{equation}
which agrees with the definition (\ref{v}). Meanwhile, the homothety
condition (\ref{homo}) is equivalent to 
\begin{equation}
\pounds _{V}E_{M}^{a}=E_{M}^{a}
\end{equation}
where $\pounds _{V}E_{M}^{a}=V^{N}D_{N}E_{M}^{a}+\partial
_{M}V^{N}E_{N}^{a}. $ This is also satisfied automatically for the geometry
constructed above in terms of $V^{a}$ and $\omega _{M}^{ab}$ as follows 
\begin{eqnarray}
\pounds _{V}E_{M}^{a} &=&V^{N}D_{N}E_{M}^{a}+\partial
_{M}V^{N}E_{N}^{a}=V^{N}T_{NM}^{a}+V^{N}D_{M}E_{N}^{a}+\partial
_{M}V^{N}E_{N}^{a} \\
&=&V^{N}T_{NM}^{a}+D_{M}\left( V^{N}E_{N}^{a}\right)
=V^{N}T_{NM}^{a}+D_{M}V^{a} \\
&=&E_{M}^{a}
\end{eqnarray}
where we have used the orthogonality of $V$ to the curvature or torsion.
Related equations appear in \cite{vasilev}, but our approach provides a OSp$%
\left( n|2\right) $ gauge symmetry basis for introducing Eq.(\ref{vielbein})
and the rest of the geometrical equations. Also, a similar problem was
discussed in \cite{strominger} in a less geometrical formalism and in the
absence of the gauge field $A_{M}.$ In our case we are interested in
solutions of the equations that permit the imposition of the constraints $%
Q_{ij}\sim Q_{i\alpha }\sim Q_{\alpha \beta }\sim 0.$

The geometry described by $E_{M}^{a}$ is fully determined by the functions $%
\omega _{M}^{ab}\left( X\right) $ and $V^{a}\left( X\right) $ which are
constrained only by the transversality condition $V^{M}R_{MN}^{ab}=0,$ but
are otherwise arbitrary. The solution space includes the most general
gravitational metric in $d$ dimensions as already seen in the previous
section. The formalism in this section provides a more covariant solution
and permits the construction of the general interacting two-time physics for
spinning particles.

\section{Conclusion and discussion}

The choice of coordinates $\kappa ,w,x^{\mu }$ and solution of background
fields used above emphasizes a basis that is convenient for deriving the
free massless relativistic particle from two time physics in the case of
zero background fields. In this basis it was easy to eliminate one timelike
and one spacelike coordinates through the gauge choices $\kappa \left( \tau
\right) =1,$ $p_{w}=0,$ leaving the usual timelike coordinate as a component
of the $d$-dimensional vector $x^{\mu }\left( \tau \right) .$ With this
choice of time we interpreted the theory and the background fields, as
discussed above. However, as we have already seen in the flat case, other
choices of the time coordinate produce very different physical
interpretations from the point of view of the one-time observer, even though
the two time physics theory is the same. In the general theory it is also
possible to work in other coordinates that are convenient to solve the Sp$%
\left( 2,R\right) $ constraints in other Sp$\left( 2,R\right) $ gauges. Then
the choice of ``time'' embedded in the two-time theory is different.

It follows that the {\it same background fields} given above would give rise
to very different interpretation of the dynamics in one-time physics in
different Sp$\left( 2,R\right) $ gauges. For example, in the flat spinless
case, with $\gamma =g_{\mu \nu }=W^{\mu }=A_{\mu }=u=0,$ different Sp$\left(
2,R\right) $ gauges produced a class of related one-time dynamics that
included the free massless relativistic particle, the free massive
relativistic particle, the free massive non-relativistic particle, the
H-atom, the harmonic oscillator in one less dimension, the particle in AdS$%
_{d-k}\times $S$^{k}$ backgrounds for any $k=0,1,\cdots ,d-2,$ etc. In a
similar way, in the general theory all possible choices of time define a
class of one-time dynamical theories related to the same two-time dynamics
with a {\it fixed set of background fields}. Changing the background fields
changes the class of related one-time dynamical models.

In the flat case the global symmetry was SO$\left( d,2\right) .$ In the
general case the Killing symmetries of the background fields (which is
embedded in the general coordinate and gauge transformations) replaces the
global SO$\left( d,2\right) $ symmetry. The global symmetries should be
realized in the same representation for all of the different one-time
dynamical models in the same class derived from the same two-time physics
theory.

\section{Acknowledgements}

I thank E. Witten for a discussion on the homothety conditions.

\end{document}